\documentclass{elsart}
\usepackage{amssymb}

\begin{document}
\journal{Physics Letters A}
 
\begin{frontmatter}
 
\title{Periodic orbit quantization of chaotic systems with strong pruning}
\author{Kirsten Weibert, J\"org Main, and G\"unter Wunner}
\address{Institut f\"ur Theoretische Physik 1,
         Universit\"at Stuttgart, D-70550 Stuttgart, Germany}
\maketitle
 
\begin{abstract}
The three-disk system, which for many years has served as a paradigm
for the usefulness of cycle expansion methods, represents an extremely hard
problem to semiclassical quantization when the disks are moved closer and
closer together, since (1) pruning of orbits sets in, rendering the symbolic
code incomplete, and (2) the number of orbits necessary to obtain accurate
semiclassical eigenvalues proliferates exponentially.
In this note we show that an alternative method, viz.\ harmonic inversion,
which does not rely on the existence of complete symbolic dynamics or other
specific properties of systems, provides a key to solving the problem of
semiclassical quantization of systems with strong pruning.
For the closed three-disk system we demonstrate how harmonic inversion,
augmented by a signal cross-correlation technique, allows one to
semiclassically calculate the energies up to the 28th excited state with
high accuracy.
\end{abstract}
 
\end{frontmatter}

\section{Introduction}
The question as to the connection between quantum mechanics
and classical dynamics lies at the heart of physics, and therefore 
has attracted continual attention for several decades.
A milestone for  understanding this relation in  chaotic systems was 
Gutzwiller's trace formula \cite{Gut90}, which provided the semiclassical
approximation to the quantum density of states in terms of a sum over 
all periodic orbits of the corresponding classical system.
However, it is a well known, and fundamental, problem of the trace formula
that it does not normally converge in the physical energy domain, mainly 
as a consequence of the rapid proliferation of periodic orbits with growing
period. 
Various techniques have been designed to overcome this problem, though most 
of them rely on special properties of the individual systems in point, such as
ergodicity and the existence of a complete symbolic dynamics
\cite{Cvi89,Aur92,Ber90}.
As an alternative, the method of harmonic inversion has been developed and 
proven \cite{Mai97b,Mai99a,Mai99b} in recent years to be universal in scope
as regards the extraction of semiclassical eigenvalues from trace formulae,
in the sense that (1) it can be applied  to both chaotic and regular systems 
and (2) does not require any specific properties of the system. 

It is the purpose of this Letter to show that harmonic inversion
also provides a key to solving the problem of semiclassical quantization 
of systems with symbolic dynamics that exhibit strong pruning.
As a prototype of such systems, we choose the three-disk billiard,
and demonstrate that even in the particularly challenging limit of 
mutually touching disks the method allows us to semiclassically calculate 
the energy eigenvalues up to the 28th excited state with high accuracy.  

The three-disk billiard system consists of three equally spaced hard disks 
of unit radius. 
Large distances $d$ between the centers of the disks, especially $d=6$, 
have been studied as a test case for periodic orbit quantization of chaotic 
systems in a number of investigations in recent years.
In particular, the system  has served as a model  for the utility
of the cycle expansion method \cite{Cvi89,Eck93,Eck95,Wir99}.
At large distances between the disks, the dynamics of the system is 
completely hyperbolic, and it is possible to uniquely label the orbits
by a complete symbolic code.
Also, at large distances, the basic assumption of the cycle expansion, 
namely, that the contributions of long periodic orbits are shadowed by 
contributions of corresponding combinations of short orbits, is very well 
fulfilled.
The actions and stability coefficients of the orbits are essentially 
determined by the cycle length, and the number of orbits up to a given 
action is relatively small.
The situation, however, changes completely when the disk separation is reduced:
As for decreasing $d$ all orbits become shorter, the total number of orbits 
up to a given value of the action increases rapidly.
Moreover, the parameters of the orbits turn out to be no longer simply 
determined by the cycle length, the condition of shadowing is more and more 
badly fulfilled, and, finally, at $d=2.04821419$, pruning sets 
in \cite{Han93}, i.e., part of the periodic orbits becomes unphysical in 
that they run {\em through} one of the disks, and have to be discarded, 
rendering the symbolic code incomplete.
The number of pruned orbits grows rapidly, as the   distance between 
the disks is further decreased. Finally,  
in the limiting case of touching disks, $d=2$, the system exhibits strong 
pruning, and the conditions for the convergence of the cycle expansion are 
no longer satisfied at all.

A remarkable first step towards semiclassical quantization of the closed 
three-disk system was achieved by combining the conventional cycle expansion 
with a functional equation \cite{Tan91}.
Based on periodic orbit corrections to the mean density of states,
approximations to the lowest eigenvalues were obtained from a small
set of periodic orbits up to cycle length $l=3$, for which the symbolic 
dynamics is still complete (including formally a zero length orbit).

It is our aim in this Letter to dispose of special additions to semiclassical 
quantization, like functional equations, and apply ``pure'' periodic orbit 
quantization by including longer orbits even in regions where strong pruning 
occurs.
The method applied here to the particular example of the closed three-disk
billiard will be the key for the semiclassical quantization of a large 
variety of other challenging systems lacking special structural information
on the quantum or classical level such as the existence of a complete
symbolic dynamics or the applicability of a functional equation.

\section{Periodic orbits of the closed three-disk system}
\begin{figure}
\vspace{6.5cm}
\includegraphics{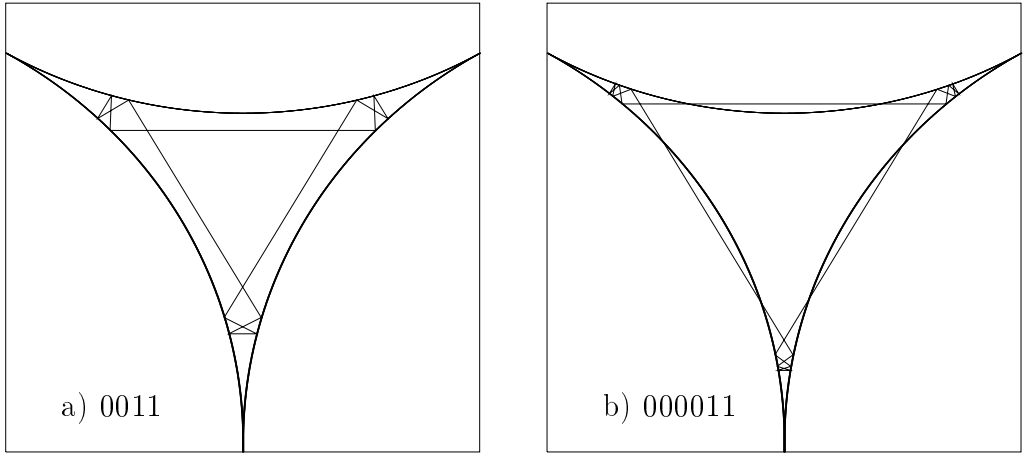}
\caption{\label{fig1} 
Two examples of periodic orbits in the closed three-disk system
(full domain representation). Orbit a) is physical, while orbit b)
runs through the disks and therefore has to be  pruned. For both orbits, 
the symmetry reduced symbolic code is given.
}
\end{figure}

In the numerical search for periodic orbits, we can make use of a symmetry 
reduction characteristic of this system \cite{Cvi89}, i.e., periodic orbits 
are calculated in a fundamental domain and labeled by a binary code with the 
alphabet '0' and '1'.
In the case of touching disks, the orbits can be grouped in ``channels''  
with the same ``tail'' (end letters) but increasing number of leading zeros 
in the symbolic code.
These orbits have the same basic geometric shape but run deeper and deeper 
into the corner formed by two touching disks, bouncing back and forth between 
the two disks, until finally -- with only two exceptions -- the channel 
breaks off because the orbits become pruned.
An example is shown in Fig.~\ref{fig1}.
Adding a leading zero to the code increases the action of the corresponding
orbit only slightly.
This means that there is a huge number of orbits with very long symbol 
lengths but relatively small, similar, actions.

\begin{figure}
\vspace{9.5cm}
\includegraphics{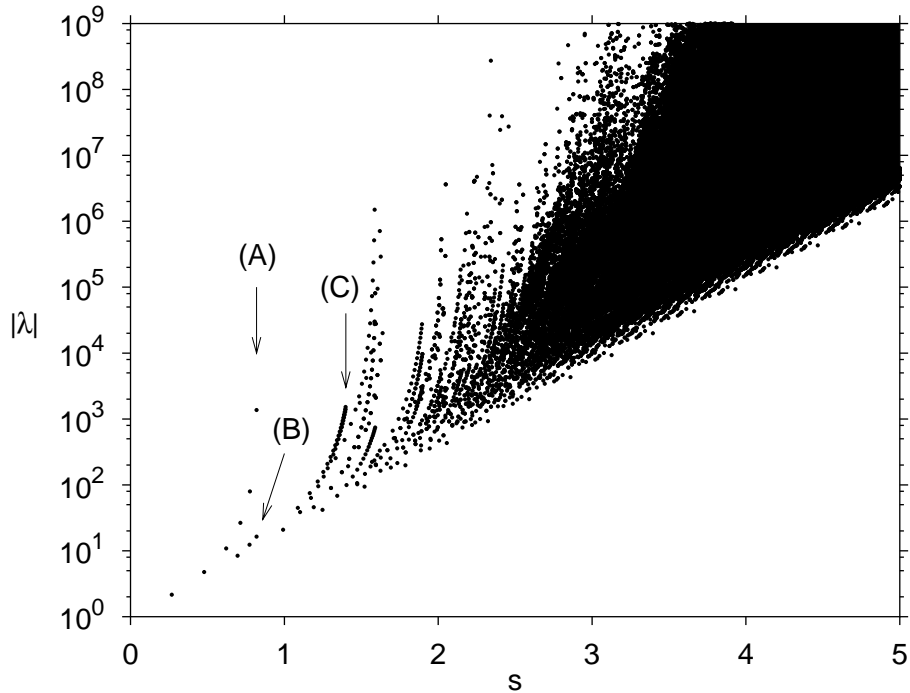}
\caption{\label{fig2} 
Distribution of periodic orbit parameters of the closed three-disk system.
The quantities plotted are the absolute value of the larger stability 
eigenvalue $|\lambda|$ versus the physical length $s$ of the orbits.
For clarity, a few channels of orbits are marked by arrows.
Channels (A) and (B) break off because of pruning.
The two channels marked (C) in fact contain infinitely many orbits, but have 
been cut off by restricting the maximum number of successive zeros in 
the symbolic code (see text).
}
\end{figure}

We searched for periodic orbits of the closed three-disk system with physical
length $s<s_{\rm max}=5.0$. The number of orbits proliferates very rapidly 
with increasing length.
Therefore, it is impossible in practice to determine all orbits up to this 
length, but one has to introduce reasonable restrictions for the orbits to 
be included.
The orbits were calculated channel by channel, starting from the shortest 
orbit in the channel and adding more and more leading zeros to the code.
As the orbits become more unstable with increasing length of the symbolic 
code, the main contributions to Gutzwiller's trace formula arise from the 
shortest orbits of each channel.
Therefore, we stopped the calculation of orbits in a given channel when the 
absolute value of the larger eigenvalue $\lambda$ of the monodromy matrix 
exceeded a set value of $\lambda_{\rm max}=10^9$.
Furthermore, to make the calculations feasible, the maximum number of 
successive zeros in the code was restricted to 20 for orbits with length 
$s<3.1$, and to 12 for orbits with $3.1<s<5.0$.
The final set of orbits obtained by an extensive numerical periodic orbit
search consists of about 5 million primitive periodic orbits, and is 
illustrated in Fig.~\ref{fig2}, where each dot in the diagram marks the 
parameter pair $(s,|\lambda|)$ of a primitive periodic orbit.
In the region $s\lesssim 2.5$ various channels are clearly distinguishable, 
a few of which are indicated by arrows. 
As mentioned above, most of the channels break off because of pruning.
However, the number of orbits grouped in a channel can be very large or, 
for the two channels labeled (C), even infinite.
[The cutoff for (C) in Fig.~\ref{fig2} is caused by restricting the number 
of successive zeros in the symbolic code as explained above.]
Note that the total number of orbits in those channels which break off
because of the restriction of the number of successive zeros in the 
symbolic code is usually much larger than in those channels where the 
number of orbits is physically restricted by pruning, which means that
pruning is the dominating effect for the cutoff of orbits in the various 
channels.

\section{Periodic orbit quantization by harmonic inversion}
The  task, and challenge, now is to extract semiclassical eigenvalues in 
a numerically stable way from the huge set of periodic orbits depicted 
in Fig.~\ref{fig2}.
To this end, we resort to the harmonic inversion technique of cross-correlated
periodic orbit sums \cite{Mai99b}, which is an extension of periodic orbit 
quantization by harmonic inversion introduced in \cite{Mai97b}.
Here, we only briefly review the basic ideas and refer the reader to the
literature for details.
For simplicity, though without loss of generality, we focus on billiard 
systems, where the shape of the orbits is independent of the energy  
$E=\hbar^2k^2/2m$, with $k$ the wave number, and the classical action of
orbits reads $S=\hbar ks$, with $s$ the physical length of the orbit.

The starting point is to introduce a weighted density of states in terms of $k$
\begin{equation}
   \varrho_{\alpha\alpha'}(k)
 = -{1\over\pi} \, {\rm Im} \, g_{\alpha\alpha'}(k) \; ,
\end{equation}
with
\begin{equation}
   g_{\alpha\alpha'}(k)
 = \sum_n {b_{\alpha n}b_{\alpha' n} \over k-k_n+{\rm i}\epsilon}\ ,
\label{g_ab_qm}
\end{equation}
where $k_n$ is the wave number eigenvalue of eigenstate $|n\rangle$
and
\begin{equation}
 b_{\alpha n} = \langle n|\hat A_\alpha|n\rangle
\end{equation}
are the diagonal matrix elements of a chosen set of $N$ linearly independent
operators $\hat A_\alpha$, $\alpha=1,2,\dots, N$.
The Fourier transform of (\ref{g_ab_qm}) yields an $N\times N$
cross-correlated signal
\begin{equation}
 C_{\alpha\alpha'}(s) = {1\over 2\pi}\int_{-\infty}^{+\infty}
      g_{\alpha\alpha'}(k){\rm e}^{-{\rm i}sk}{\rm d}k  
 = -{\rm i} \sum_n b_{\alpha n}b_{\alpha' n} {\rm e}^{-{\rm i}k_ns} \; .
\label{C_ab_qm}
\end{equation}
A semiclassical approximation to the cross-correlated signal (\ref{C_ab_qm})
has been derived in \cite{Mai99c,Hor00}.
The cross-correlated periodic orbit signal reads
\begin{equation}
   C_{\alpha\alpha'}^{\rm sc}(s)
 = -{\rm i}\sum_{\rm po} {a_{\alpha,{\rm po}}\, a_{\alpha',{\rm po}}\,
    s_{\rm po}\, {\rm e}^{-{\rm i}{\pi\over 2}\mu_{\rm po}}
    \over r |\det(M_{\rm po}-1)|^{1/2}}
    \delta\left(s-s_{\rm po}\right) \, ,
\label{C_ab_sc}
\end{equation}
where $r$ is the repetition number counting the traversals of the primitive 
orbit, and $M_{\rm po}$ and $\mu_{\rm po}$ are the monodromy matrix and 
Maslov index of the orbit, respectively.
The weight factors $a_{\alpha,{\rm po}}$ are classical averages over
the periodic orbits
\begin{equation}
 a_{\alpha,{\rm po}} = {1\over s_{\rm po}} \int_0^{s_{\rm po}}
  A_\alpha({\bf q}(s),{\bf p}(s)) {\rm d}s \; ,
\label{a_po}
\end{equation}
with $A_\alpha({\bf q},{\bf p})$ the Wigner transform of the operator
$\hat A_\alpha$.
Semiclassical approximations to the eigenvalues $k_n$ and eventually also
to the diagonal matrix elements $\langle n|\hat A_\alpha|n\rangle$ are
now obtained by adjusting the semiclassical cross-correlated periodic orbit
signal (\ref{C_ab_sc}) to the functional form of the quantum signal
(\ref{C_ab_qm}).
The numerical tool for this procedure is an extension of the harmonic 
inversion method to the signal processing of cross-correlation functions 
\cite{Nar97,Man98}.

The resolution of the results depends on the signal length.
For  a one-dimen\-sion\-al signal, the method requires a signal length of 
$s_{\rm max}\approx 4\pi\bar\rho(k)$, with $\bar\rho(k)$ the average density
of states to resolve the eigenvalues \cite{Mai97b}.
This means that all periodic orbits up to the scaled action $s_{\rm max}$ 
have to be included.
The advantage of using the cross-correlation approach is based on the 
realization that the total amount of independent information contained in
the $N\times N$ signal is $N(N+1)$ multiplied by the length of the signal, 
while the total number of unknowns (here $b_{\alpha n}$ and $k_n$) is $(N+1)$
times the total number of poles $k_n$. 
Therefore the informational content of the $N\times N$ signal per unknown 
parameter is increased (as compared to the one-dimensional signal) by 
roughly a factor of $N$, and the cross-correlation approach should lead to 
a significant improvement of the resolution.
The technique has been  successfully applied in the periodic
orbit quantization of an integrable system \cite{Mai99b}. We
will now demonstrate  the power of the method in the semiclassical
quantization of the (chaotic) closed three-disk system.
As in Ref.~\cite{Tan91}, we concentrate on the quantum states with $A_1$ 
symmetry of the $C_{3v}$ group.

\begin{figure}
\vspace{9.2cm}
\includegraphics{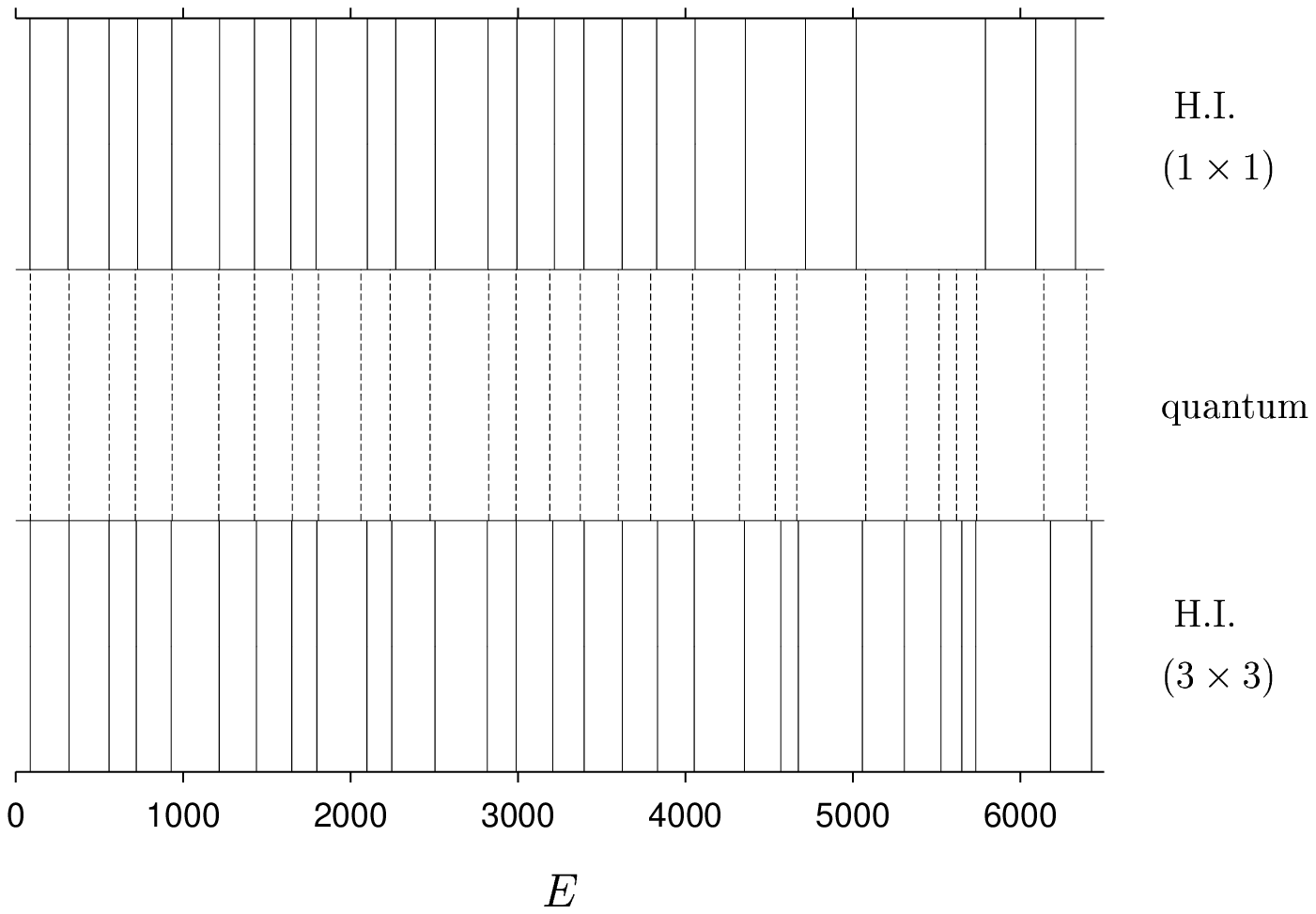}
\caption{\label{fig3} 
Energy eigenvalues of the closed three-disk system. Top spectrum:
Harmonic inversion of a one-dimensional periodic orbit signal of length 
$s_{\rm max}=4.9$.
Middle spectrum: Exact quantum states (from \protect\cite{Sche91}).
Bottom spectrum: Semiclassical eigenvalues 
obtained by harmonic inversion of a $3\times 3$ cross-correlated 
periodic orbit signal of length $s_{\rm max}=4.8$.
The operators used to build  the signal were ${\bf 1}$ (unity), $r^4$, 
and $L^4$. The 29 lowest eigenvalues are well reproduced by 
semiclassical quantization.
}
\end{figure}

The semiclassical cross-correlation signal (\ref{C_ab_sc}) is constructed
using the set of about 5 million periodic orbits represented in 
Fig.~\ref{fig2}.
We first construct a one-dimensional signal ($N=1$) by simply choosing 
$\hat A_1={\bf 1}$, i.e., the unity operator.
Fig.~\ref{fig3} contains  the results for the eigenvalues $k_n$ of the closed
three-disk system in terms of the corresponding energy values $E_n=k_n^2/2$.
The uppermost spectrum is the result from  harmonic inversion  obtained from 
the one-dimensional signal of length $s_{\rm max}=4.9$. 
The spectrum in the middle shows the exact quantum results, taken from
Ref.\ \cite{Sche91}, for comparison.
It can be seen that the  lowest 20 eigenvalues are well reproduced by the
results of the harmonic inversion of the $1 \times 1$  signal.
The major part of the small deviations is  due to the error inherent in
the semiclassical approximation.
However, it can also be seen that the signal length is not sufficient 
to resolve the higher eigenvalues, with energies $E\gtrsim 4500$,
where the mean density of states
with respect to $k$ (which grows $\sim\sqrt{E}$) is too large.

A further increase of the signal length in an attempt to resolve more 
eigenvalues is clearly inhibited by the rapid proliferation of periodic 
orbits in this system (cf. Fig.~\ref{fig2}).
Instead, we now exploit the cross-correlation technique to 
significantly improve the resolution. We   construct the cross-correlated 
periodic orbit signal (\ref{C_ab_sc}) using a set of three independent 
operators, viz.\ the operators $\hat A_1={\bf 1}$ (unity), $\hat A_2=r^4$, 
and $\hat A_3=L^4$, with $r$ and $L$ the distance from the center of the 
billiard and the angular momentum, respectively. The bottom spectrum
in Fig.~\ref{fig3} shows the results for the eigenvalues of the closed 
three-disk system obtained from the harmonic inversion of the $3\times 3$ 
set of cross-correlated periodic orbit signals.
Only the converged frequencies are presented, which have been selected by the
criterion that their amplitude (multiplicity) should be close to 1 and the 
imaginary part of the frequencies close to zero (as the eigenvalues must 
be real).
With the cross-correlated signal, all eigenvalues up to $E\lesssim 6500$ 
(a total of 29 levels) can be  well reproduced. 

The accuracy of the semiclassical results presented in Fig.~\ref{fig3} 
may at first glance seem  surprising when one recalls the fact that 
the set of periodic orbits (see Fig.~\ref{fig2}) used for constructing 
the signal (\ref{C_ab_sc}), although being huge, is not strictly complete.
For example, the channel marked (C) in Fig.~\ref{fig2} was cut off
by restricting the number of successive leading zeros in the symbolic code.
However, the Maslov indices of adjacent orbits in that channel differ by 
2, and therefore the periodic orbit amplitudes alternate in sign, which
leads to approximate cancellations of terms in the periodic orbit signal.
Furthermore, very unstable orbits with stability eigenvalues
$|\lambda|>\lambda_{\rm max}=10^9$ were ignored in the analysis.
By lowering the limit $\lambda_{\rm max}$ we have checked that indeed
the very unstable orbits do not carry important information about the
semiclassical spectrum, i.e., the results change only insignificantly.
On the other hand, increasing $\lambda_{\rm max}$, and thus the total number 
of orbits used for harmonic inversion, will still slightly improve the 
accuracy of the semiclassical spectrum.

In conclusion, we have successfully carried out a ``pure'' periodic orbit 
quantization of a chaotic system with extremely rapid proliferation and 
strong pruning of orbits.
We have demonstrated that harmonic inversion of cross-correlation signals
is indeed a powerful method for periodic orbit quantization, allowing the 
numerically stable handling even of huge periodic orbit sets.
It should also be stressed that the semiclassical eigenvalues were
obtained without having to resort to the mean density of states, which
is difficult to calculate, including all necessary correction terms, even 
in the case of the closed three-disk system \cite{Sche91}.
The present results may stimulate future work on other challenging systems
without special structural information on the quantum or classical level.
One example with strong pruning of the symbolic dynamics is the 
three-dimensional generalization of the closed three-disk billiard, i.e., 
a system consisting of four touching spheres at the corners of a regular 
tetrahedron.
By contrast to the closed three-disk system this is an open system where
no functional equation can be applied.

\section*{Acknowledgments}
We thank A.~Wirzba for communicating to us quantum mechanical data 
of the closed three-disk system.
This work was supported by the Deutsche For\-schungs\-ge\-mein\-schaft and
Deutscher Akademischer Austauschdienst.


\end{document}